\documentclass[iop]{emulateapj}
\usepackage{mathptmx}
\usepackage{fancyvrb}
\usepackage{epstopdf}

\usepackage{amsmath}
\usepackage{amssymb}

\usepackage{graphicx}
\usepackage{float}
\usepackage{natbib}
\usepackage{footnote}
\usepackage{bbold}
\usepackage{enumerate}
\usepackage{paralist}

\DefineVerbatimEnvironment{code}{Verbatim}{fontsize=\small}
\DefineVerbatimEnvironment{example}{Verbatim}{fontsize=\small}

\def\gtorder{\mathrel{\raise.3ex\hbox{$>$}\mkern-14mu
             \lower0.6ex\hbox{$\sim$}}}
\def\ltorder{\mathrel{\raise.3ex\hbox{$<$}\mkern-14mu
             \lower0.6ex\hbox{$\sim$}}}

\submitted{Submitted to ApJ}

\shorttitle{Optimal coaddition of images}

\shortauthors{Zackay \& Ofek}

\begin{document}




\title{How to coadd images? II. A coaddition image that is optimal for any purpose in the background dominated noise limit}
\author{Barak Zackay\altaffilmark{1} and
Eran O. Ofek\altaffilmark{1}
}
\altaffiltext{1}{Department of Particle Physics and Astrophysics, Weizmann Institute
  of Science, 76100 Rehovot, Israel}

\begin{abstract}
Image coaddition is one of the most basic operations
that astronomers perform.
In Paper~I, we presented the optimal ways to coadd images in order to detect faint sources and to perfrom flux measurements under the assumption
that the noise is approximately Gaussian.
Here, we build on these results and derive from first principles a coaddition
technique which is optimal for any hypothesis testing
and measurement (e.g., source detection, flux or shape measurements and star/galaxy separation),
in the background-noise-dominated case.
This method has several important properties.
The pixels of the resulting coadd image are uncorrelated.
This image preserves all the information (from the original
individual images) on all spatial frequencies.
Any hypothesis testing or measurement that can be done
on all the individual images simultaneously,
can be done on the coadded image without any loss
of information.
The PSF of this image is typically as narrow, or narrower
than the PSF of the best image in the ensemble.
Moreover, this image is practically indistinguishable from a regular single image, meaning that any code that measures any property on a regular astronomical image can be applied to it unchanged.
In particular, the optimal source detection statistic derived in paper~I is reproduced by matched filtering this image with its own PSF.
This coaddition process, which we call proper coaddition, can be understood
as a the maximum signal-to-noise ratio
measurement of the Fourier transform of the image, weighted
in such a way that the noise in the entire Fourier domain is 
of equal variance.
This method has important implications for multi-epoch
seeing-limited deep surveys, weak lensing galaxy shape measurements,
and diffraction-limited imaging via speckle observations.
The last topic will be covered in depth in future papers.
We provide an implementation of this algorithm in MATLAB.
%
%
%
%
%
\end{abstract}


\section{Introduction}
\label{sec:Introduction}
%
%
Practically all ground-based surveys are subject to the ever-changing atmospheric seeing conditions, sky brightness and transparency.
These changes produce large variations in both resolution and depth of astronomical observations. It is often necessary to image the same area of sky repeatedly, in order to both increase survey depth, and to allow time-dependent science cases. Examples for such surveys are LSST \citep{LSST}, DES \citep{DES}, PTF \citep{PTF}, CFHTLS \citep{CFHTLS}, Pan-STARRS \citep{PANSTARRS}, SkyMapper \citep{SKYMAPPER} and SDSS \citep{SDSS}.

In order to simultaneously pursue all science cases that require coaddition, there is a demand for a method that preserves all the information that exists in the data in one coadded image, leaving the original data redundant. 
Putting this in formal language, we would like to construct an image which is a sufficient statistic\footnote{A statistic is sufficient with respect to a model and its associated parameter if no other statistic that can be calculated from the same sample provides any additional information as to the value of the models parameter.} for any further processing of the data, such as source detection, applying shear measurements or performing astrometric measurements.

It is non-trivial that such an image exists, as currently no coaddition technique is used for all purposes and existing methods fail to encompass all the data.
For example, for many applications it is customary to use only images with favorable seeing.
In some cases (e.g., lucky imaging; \citealp{Law06})
it is common to discard $>90\%$ of the data. Even for the simple purpose of maximal sensitivity for source detection, various authors use different weighing
schemes for the image combination (e.g., \citealp{Annis2014,Jiang2014}).

This paper is number II in a series of papers on the problem of image coaddition. In paper I \citep{Coad1} we presented the best way to coadd images for source detection and flux measurement under several different regimes of noise properties. In this paper, we focus on the background-noise-dominated regime and describe a coaddition method that is optimal for performing any measurement or hypotheses testing on the data (for non-variable sources). The product of this coaddition method is therefore a sufficient statistic for any further processing.
We note that the background dominated noise limit is the most common case in astronomical observations.

We show that the coadded image also has an intuitive explanation as being the best ``proper'' image, where we define an image being proper if its additive noise component satisfies that
\begin{inparaenum}[\itshape i\upshape)]
\item All pixels have equal variance and
\item All pairs of pixels are uncorrelated.
\end{inparaenum}
We also show that this image behaves exactly like an ordinary image, which means that any single-image analysis code that exists will operate correctly on it, exploiting all the information that existed in the original dataset.

After completing the work on this paper, we became aware that a similar method to the one we derive was already presented in \cite{Homrighausen} without any formal derivation.
They refer to this method, as well as another deconvolution method, as a private communication from N. Kaiser (2004).
\cite{Homrighausen} mention that this
method has some associated optimality properties,
but they argue that the gain from using this method is marginal.
Here we show that the method presented in this paper is
the optimal way to coadd a stack of images in the background-dominated-noise
limit, and a sufficient statistic for any further signal processing.
Tested on real data, we demonstrate that
for seeing-limited surveys this method provide a few percents to 25\% increase in survey speed
compared with weighted coaddition schemes (e.g., Annis et al. 2014; Jiang et al. 2014).
Furthermore, we claim that usage of this method may have additional far-reaching consequences, as we show in this and future papers.
Homrighausen et al. (2011) further argue
that this method is
a biased estimator of the {\it true} image and that
another method - namely the 
deconvolution method, although numerically unstable, 
has the smallest variance
of all methods.
It is worth while to analyze these statements as they may
help us understand the apparent conflict between our results
and those of Homrighausen et al. (2011).
In practice a reconstruction of the true image is impossible (unless the $S/N$ is infinite and the PSF is known exactly) as in general our knowledge of the spatial high frequencies is greatly suppressed by the PSF.
Therefore, the desired property from a coaddition method is not to reconstruct the true image,
but rather to provide a sufficient statistic representing all the observed data in one image. This allows us to test hypotheses and to measure properties of sources in the true image, in an unbiased way and with the highest precision.
The metrics which Homrighausen et al. (2011) apply to compare different image coaddition techniques are image quality metrics like the PSF width and the mean integrated squared error of the image from the true scene. These metrics do not represent accurately the requirement for maximal sensitivity of measurements made using the coadded image.
These are the major differences between the approach taken by
Homrighausen et al. (2011) and our methodology.

The structure of the paper is as follows:
In \S\ref{sec:StatisticalBackground} we present the statistical model and assumptions.
In \S\ref{sec:properImage} we present an intuitive definition of what is a natural astronomical image, which we call a {\it proper image}.
In \S\ref{sec:BestProperImage} we convert the optimal source detection statistic presented in \cite{Coad1} to a proper image.
In \S\ref{sec:Sufficiency} we show that the derived image is a sufficient statistic for any statistical test or decision based on the data.
In \S\ref{sec:ExamplesVisualization} we show examples of the application of the proper image coaddition statistic to both simulated PSFs and real data.
In \S\ref{sec:Code} we discuss the implementation details relevant to the successful application of the coaddition method.
In \S\ref{sec:Summary} we highlight the key results and equations, and discuss the implications of this coaddition method to future surveys.

\section{Statistical formalism}

\label{sec:StatisticalBackground}

Following the same notation as in paper I, we denote the $j$'th background subtracted image, in a series of observations of the same position, by $M_j$.
All the images are assumed to be registered. 
This image can be described by the convolution of the true, background subtracted, noiseless image $T$ with the image point spread function (PSF) $P_j$, added to which is a noise term $\epsilon_{j}$,


\begin{align}
M_j = (F_jT)\otimes P_j + \epsilon_j\, ,
\end{align}
where $F_j$ is a multiplicative factor that describes the transparency\footnote{More accurately, the product of transparency, exposure time, effective collecting area of the optical system and the detector efficiency.} and $\otimes$ represents convolution.
The noise term $\epsilon_j$ has a variance:
\begin{align} V[\epsilon_j] = B_j + (F_jT)\otimes P_j\,,\end{align}
where $B_j$ is the image background (or more generally, the total variance that is homogeneous across the image -- i.e., including the read noise).
Here, for simplicity, we assume that the image gain is $1$, and that the background noise is the dominant source of noise, i.e \begin{align}(F_jT)\otimes P_j << B_j\, .\end{align}
Using this, we can write \begin{align} \label{Eq.BackgroundDominated}V[\epsilon_j]  \equiv \sigma_j^2 \approx B_j\,.\end{align}

We further assume that the Poisson noise of the background to be Gaussian with good approximation (e.g., $B_j\gtrsim 30$), and that the noise component in different pixels is uncorrelated.

Equation \ref{Eq.BackgroundDominated} holds
in many astronomical images.
Given typical ground-based visible-light sky-background value (e.g., $m_{{\rm sky}}\sim 18-21\,{\rm mag}\,{\rm{arcsec}^{-2}}$) and typical fraction of an arcsecond pixel size.
This means that this condition is satisfied for all objects with surface brightness
fainter than about $21\,{\rm mag}\,{\rm{arcsec}^{-2}}\,.$

\subsection{Coadding images for source detection}\label{subsec:SourceDetection}

In paper I we presented the best coadded image for source detection under the assumption that the noise in all observations is (locally) spatially homogeneous, Gaussian, and that the noise component of different pixels is uncorrelated.
This image is given by:
\begin{align} \label{Eq.S}
S = \sum_j\left(\frac{F_j\overleftarrow{P_j}}{\sigma_j^2}\right)\otimes M_j\,,
\end{align}
where $\overleftarrow{P_j}(x,y) = P_j(-x,-y)$.

This image is essentially the equivalent of a matched filtered image for an ensemble of images.
Therefore $S$ is a score image, showing the log-likelihood for the existence of a source in each position. 
Matched-filtered images, while being best for source detection, are not sharp because they have an effective PSF which is twice the one of the original image. 
This is the reason that albeit optimal for source detection, astronomers usually do not look at matched-filtered images.
Furthermore, matched-filtered images cannot be intuitively interpreted for any purpose other than single source detection because neighboring pixels of these images are highly correlated. For example, performing galaxy shape measurements requires the spatial covariance matrix.

\section{Noise properties of images, and the definition of a proper image}\label{sec:properImage}

A major difference between the coaddition product described in Equation \ref{Eq.S} and a regular astronomical observation, is that the noise in neighboring pixels of $S$ is highly correlated due to the matched-filtering operation.
The correlated noise properties of matched-filtered images requires special care whenever we perform any statistical test or measurement. For example, fitting a model (e.g., a galaxy shape model)
to a light source, in an image with correlated noise, requires considerable effort in diagonalizing the covariance matrix of the many pixels involved. 
This motivates us to define a {\it proper image} to be an image that satisfies the following two conditions:
\begin{enumerate}[I]
\item All the pixels in the additive noise component have equal variance\footnote{Practically, the equal variance assumption can be relaxed to the assumption of a background which is not a strong function of spatial position. This is because all operations we apply are effectively local, and the image at a certain pixel involves only weighted summation of pixels which are up to a few PSF widths of the original position in all images.}. \label{CondEqualVar}
\item Different pixels of the additive noise  component ($\epsilon_j$) are uncorrelated. \label{CondUncorrelated}

\end{enumerate}
Since in most (background subtracted) astronomical images the noise can be assumed to be approximately Gaussian, Conditions \ref{CondEqualVar} and \ref{CondUncorrelated} imply that all pixels will be independent and identically distributed (i.i.d.).
A property of proper images is that the Fourier transform of its noise component also satisfies conditions \ref{CondEqualVar} and \ref{CondUncorrelated}.
The converse is also true, if the Fourier transform of an image satisfies conditions \ref{CondEqualVar} and \ref{CondUncorrelated}, the image itself is proper.

We note that any linear combination of images that
are directly measured from the telescope,
is (at least locally) a proper image.
A few examples of non-proper images:
\begin{itemize}
\item A matched-filtered image is not proper, since the noise in neighboring pixels is correlated.
\item A deconvolved image is not proper, as long range correlations are always present in images that are the product of deconvolution.
\item A proper image that has been ``flat fielded'' is no longer a proper image. The reason is that the variance of such images is position dependent. However, if the flat field is sufficiently smooth, it could be approximately regarded as proper because as we will show, all the operations done on proper coadd images are local operations.
\item An image with a strong source, that either saturates the detector, or is much brighter than the sky background is not proper. Note that any slice of this image that does not contain significant contamination from the bright source can be regarded as a proper image.
\end{itemize}

\section{The best proper image}\label{sec:BestProperImage}
There are many ways to coadd proper images
and to produce a new (coadded) proper image.
However, we are interested in a proper image
which maximizes the signal to noise ratio ($S/N$) of the faintest sources. Such an image must preserve the depth of the original set of images.
Moreover we would like our image to preserve all spatial information, which could be interpreted as having the best possible $S/N$ in each spatial frequency. As we will show below, these two requirements coincide.
The best starting point for such an image is to start by turning the statistic $S$ (Eq. \ref{Eq.S}), which is optimal for source detection, into a proper image.
In order to make a proper image from the statistic $S$, we examine its Fourier transform:
\begin{align}\label{Eq.Shat}
\widehat{S}(f) = \sum_{j}\frac{F_j}{\sigma_j^2}\overline{\widehat{P_j}(f)}\widehat{M_j}(f)\,.
\end{align}
Here, $f = (f_1,f_2)$ is the spatial frequency (two dimensional) index, and we will omit it from now on.
The hat sign $\widehat{\;\;}$ represents Fourier transform,
while the accent $^{-}$ marks complex conjugation.
When the overline accent appears above the hat symbol it means that the complex conjugate follows the Fourier transform.

The additive noise frequencies in $\widehat{S}$ are uncorrelated. They were uncorrelated in each of the Fourier transformed original images $\widehat{M_j}$, and therefore are uncorrelated after their weighted addition. Furthermore, the frequencies of $\widehat{M_j}$ have equal variance as we assumed the images $\widehat{M_j}$ are proper (this happens whenever the source noise is negligible).
However, the frequencies of the product $\overline{\widehat{P_j}}\widehat{M_j}$ are not of equal variance, as $|\widehat{P_j}|$ can change drastically between low and high frequencies. In order to transform this image into a proper image, all we need to do is to normalize each frequency by its own standard deviation.
Calculating the standard deviation of each pixel, we can get the Fourier transform of the proper image:

\begin{align}\label{eq:Rhat}
\widehat{R} = \frac{\sum_{j}\frac{F_j}{\sigma_j^2}\overline{\widehat{P_j}}\widehat{M_j}}{\sqrt{\sum_{j}\frac{F_j^2}{\sigma_j^2}|\widehat{P_j}|^2}}\,.
\end{align}
The noise in $\widehat{R}$ satisfies conditions \ref{CondEqualVar} and \ref{CondUncorrelated}, and hence, its inverse Fourier transform $R$ is also a proper image.

The reason that $R$ preserves all spatial information, is because in the computation of each frequency of $\widehat{R}$, we add random variables scaled by their (conjugate) expectation, divided by the variance. We can identify this operation as the maximal $S/N$ addition of random variables, shown in the Appendix of paper I.
This means that each of the frequencies of $\widehat{R}$ is the maximum $S/N$ linear estimator of $\widehat{T}$, where each frequency is scaled to have the same noise variance.
Therefore, $R$ is not just a proper image, it is the optimal proper image in the limit of background dominated noise.

Interestingly, as $R$ is a proper image, and behaves like a standard image, a natural question is therefore what is its PSF?
The PSF of an image is the normalized response of the image to a point-source located at position $(0,0)$, i.e., when $T = \delta_{(0,0)},\epsilon_j = 0$, (where $\delta(0,0)$ denotes an image with one at the origin and zero otherwise) which means $\widehat{T} = 1, \widehat{\epsilon_j} = 0$. For example, by evaluating $\widehat{M_j}$ with these substitutions, we get $\widehat{P_j}$:
\begin{align}\label{eq:SubsMbyP}
\widehat{M_j} = F_j\widehat{T}\widehat{P_j} + \widehat{\epsilon_j} = F_j\widehat{P_j}\propto \widehat{P_j}\,.
\end{align}
Repeating this process for $R$, by substituting Equation~\ref{eq:SubsMbyP} into Equation~\ref{eq:Rhat} we get the Fourier transform of the PSF of the proper coaddition image, which we denote by $P_R$:
\begin{align}\widehat{P_R} \propto \frac{\sum_{j}\frac{F_j}{\sigma_j^2}\overline{\widehat{P_j}}F_j\widehat{P_j}}{\sqrt{\sum_{j}\frac{F_j^2}{\sigma_j^2}|\widehat{P_j}|^2}} = \sqrt{\sum_{j}\frac{F_j^2}{\sigma_j^2}|\widehat{P_j}|^2}\,.  \end{align}
Normalizing $P_R$ to have unit sum, we get:
\begin{align}\label{eq:Prhat} \widehat{P_R} = \frac{\sqrt{\sum_{j}\frac{F_j^2}{\sigma_j^2}|\widehat{P_j}|^2}}{\sqrt{\sum_j\frac{F_j^2}{\sigma_j^2}}} \equiv \frac{\sqrt{\sum_{j}\frac{F_j^2}{\sigma_j^2}|\widehat{P_j}|^2}}{F_R} \,,\end{align}
where $F_R$ is the equivalent of transparency for the properly coadded image.
The standard deviation of the properly coadded image (Eq. \ref{eq:Rhat}), $\sigma_R$, is given by:
\begin{align}
\sigma_R = 1\,,
\end{align}
Finally, we can express the full statistical model for the coadded image $R$ by:
\begin{align}
R = F_RT\otimes P_R + \epsilon_R\,,
\end{align}
where $\epsilon_R$ are uncorrelated Gaussian variables with  \begin{align}V[\epsilon_R] = \sigma_R^2 = 1\,.\end{align}
 
Not surprisingly, the log-likelihood ratio for the existence of a point source in the coadded image, defined in Equation~\ref{Eq.S}, $S$, can be computed by matched-filtering $R$ with its PSF:  
\begin{align}
S = \frac{F_R}{\sigma_R^2}\overleftarrow{P_R}\otimes R\,.
\end{align}
Equivalently, in fourier space:
\begin{align}\label{eq:RelationSR}
\widehat{S} = \frac{F_R}{\sigma_R^2}\overline{\widehat{P_R}}\widehat{R}\,.
\end{align}
Therefore, for source detection, this method is identical
to the technique presented in paper~I.

An important remark is that $P_R$ is numerically stable. In contrast with deconvolution that is often suggested for coaddition, there is no division by small numbers in the process of calculating $R$ and $P_R$.
This is evident from the fact that every effective kernel of every measurement $M_j$ in the numerator of Equation~\ref{eq:Rhat} is proportional to $|P_j|$, while the denominator is larger then $|P_j|$ (triangle inequality). Hence, in the limit of $P_j\rightarrow 0$  Equation~\ref{eq:Rhat} will not diverge.

A demonstration of the properly coadded images that we get when adding several sets of simulated examples are shown in Figure \ref{fig:PSFs} (see \S \ref{sec:ExamplesVisualization}).

\section{$R, P_R$ are sufficient for any further processing}
\label{sec:Sufficiency}
Although having an image that you can intuitively understand is useful, the importance of the best proper image does not end here.
In this section, it is our goal to show that any scientific question about the data (e.g., galaxy shapes), which is not time variable, can be answered using the proper image, without any further necessity to access or even store the original data.
In \S \ref{subsec:HypoTest} we will show that $R$ and $P_R$ are sufficient for a statistical decision between any two generative hypotheses, meaning that using these will not result in any information loss with respect to performing the hypothesis testing using the entire set of images simultaneously.
In \S \ref{subsec:Measurement} we will extend this result to the case of performing any measurement (e.g., photometry, astrometry and shape measurements).

\subsection{$R, P_R$ are sufficient for any hypothesis testing task}\label{subsec:HypoTest}

Assuming that we want to decide between two generative models for $T$. In the case of the first (null) hypothesis $\mathcal{H}_0$, our model for the truth image is $T_0$, and in the case of the second hypothesis $\mathcal{H}_1$, our model for the truth image is $T_1$.
%
Given $\mathcal{H}_0$, the images are of the form:
\begin{align}\label{eq:H0}M_j = (F_jT_0)\otimes P_j + \epsilon_j,\end{align}
and given $\mathcal{H}_1$, the images are of the form:
\begin{align}\label{eq:H1}M_j = (F_jT_1)\otimes P_j + \epsilon_j\,.\end{align}
Rearranging Equation \ref{eq:H0}, and noting that convolution
with a delta function ($\delta_{(0,0)}$) returns a constant, we get:
\begin{align}M_j = (F_jT_0)\otimes P_j + \epsilon_j = (\delta_{(0,0)} F_j)\otimes (T_0\otimes P_j) + \epsilon_j,\end{align}
For convenience, we will add a reference hypothesis, $\mathcal{H}_2$ of the form (i.e., no sources)
\begin{align}M_j = \epsilon_j\,.\end{align}
Applying our solution from paper I for the point source detection problem, we can calculate the log likelihood ratio\footnote{The lemma of \cite{NeymanPearsonLemma} states that for simple hypotheses testing the log-likelihood ratio test is the most powerful test.} between hypothesis $\mathcal{H}_\alpha$ and $\mathcal{H}_2$, where $\alpha\in \{0,1\}$.
\begin{align}\mathcal{L}_{\alpha,2} = \ln(\mathcal{P}[M|\mathcal{H}_\alpha]) - \ln(\mathcal{P}[M|\mathcal{H}_2])\,,\end{align}
where $\mathcal{P}[M|\mathcal{H}_\alpha]$ denotes the probability to observe $M$ given the hypothesis $\mathcal{H}_\alpha$ is true.
In paper I we showed that this log likelihood ratio could be calculated for all positions by the statistic $S_{\alpha,2}$, given by: 
\begin{align}
\widehat{S_{(\alpha,2)}} = \sum_{j}\frac{F_j}{\sigma_j^2}\overline{\widehat{T_\alpha}\widehat{P_j}}\widehat{M_j}\,.
\end{align}
The test statistic that we are really interested in calculating for all positions is
\begin{align}
S_{(0,1)} &= \ln(\mathcal{P}[M|\mathcal{H}_0]) - \ln(\mathcal{P}[M|\mathcal{H}_1]) \\ \nonumber &=  S_{(0,2)} - S_{(1,2)}\,.
\end{align}
This is given by:
\begin{align}\label{EqHypTestingSingle}
\widehat{S_{(0,1)}} &= \sum_{j}\frac{F_j}{\sigma_j^2}(\overline{\widehat{T_0}} - \overline{\widehat{T_1}})\overline{\widehat{P_j}}\widehat{M_j}  \\ \nonumber &= (\overline{\widehat{T_0}} - \overline{\widehat{T_1}})\sum_{j}\frac{F_j}{\sigma_j^2}\overline{\widehat{P_j}}\widehat{M_j} \\ \nonumber & = (\overline{\widehat{T_0}} - \overline{\widehat{T_1}}) \widehat{S}.
\end{align}
This shows that using $\widehat{S}$ (defined in Eq. \ref{Eq.Shat}) alone we can calculate the same numerical value for the log likelihood ratio as if we had all the images. This makes it clear that all the information needed in order to make a statistical decision is cast into $S$.

In order to perform the hypothesis testing all that remains is to specify the decision boundary. For that we need to calculate the expectancy and variance of $S_{(0,1)}$ given both hypotheses.
Using Equations \ref{Eq.Shat},\ref{eq:Prhat}, \ref{eq:H0} and \ref{eq:H1}, we can calculate the expectancy of $\widehat{S}$ given $\mathcal{H}_\alpha$, which is given by:

\begin{align}E[\widehat{S}|\mathcal{H}_\alpha] = \widehat{T_\alpha}\sum_j{\frac{F_j^2}{\sigma_j^2}|\widehat{P_j}|^2} = \widehat{T_\alpha}F_R^2|\widehat{P_R}|^2\,.\end{align}
Similarly, 
\begin{align}
E[\widehat{S_{(0,1)}}|\mathcal{H}_\alpha] &= T_\alpha\overline{(\widehat{T_0}-\widehat{T_1})}F_R^2|\widehat{P_R}|^2\,.
\end{align}
To calculate the variance of $S_{(0,1)}$, we first need its description in the image plane, which is (from Eq. \ref{EqHypTestingSingle}):

\begin{align}
S_{(0,1)} = \overleftarrow{(T_0-T_1)}\otimes \left(\sum_j{\frac{F_j}{\sigma_j^2}\overleftarrow{P_j}\otimes M_j}\right) = \overleftarrow{(T_0-T_1)}\otimes S \,,
\end{align}
and therefore, the variance of $S_{(0,1)}$ is:
\begin{align}
V[S_{(0,1)}] = \sum_{j}{ \sum_x{\left(\frac{F_j}{\sigma_j^2}\overleftarrow{(T_0-T_1)\otimes P_j}\right)^2(x)V(M_j(x))}}\,. \end{align}
Using the Parseval equality, and the fact that $V[M_j] = \sigma_j^2$, we get:
\begin{align} V[S_{(0,1)}] = \sum_j{\sum_{f}{\frac{F_j^2}{\sigma_j^2}\left|\overline{(\widehat{T}_0-\widehat{T}_1)}\overline{\widehat{P}_j}\right|^2}}\,. \end{align} 
Noticing that $F_R^2|\widehat{P_r}|^2 =  \sum_j{\frac{F_j^2}{\sigma_j^2}|\overline{\widehat{P}_j}|^2}$ (Eq. \ref{eq:Prhat}), and simplifying, we get:
\begin{align} V[S_{(0,1)}] = 
\sum_{f}{|(\widehat{T}_0-\widehat{T}_1)|^2F_R^2|\widehat{P}_R|^2}\,.
\end{align}
This means that any decision between two simple generative hypotheses can be done directly on $S$.
To correctly normalize and analyze the derived score, we also need to store $F_R^2|\widehat{P_R(f)}|^2$. This means that $S$, $P_R$ and $F_R$, or equivalently, $R$, $P_R$, and $F_R$ are sufficient for hypothesis testing on background-dominated noise images.

Moreover, by using $R$, $P_R$ and $F_R$, we notice that the hypothesis testing process is exactly the same as if $R$ was a single image, $P_R$ its PSF and $F_R$ its transparency.
To see this, we can further develop
\begin{align}
S_{(0,1)} = R\otimes \left(F_R\overleftarrow{P_R}\otimes\overleftarrow{(T_0-T_1)}\right),
\end{align}
which is the exact expression for simple hypothesis testing on the single image $R$ with the PSF $P_R$ and transparency $F_R$.

Testing composite hypotheses, which are hypotheses that have intrinsic parameters (an example problem is the star-galaxy separation task in which each hypothesis has a different set of parameters) requires a much more elaborate statistical process, on which there is no widespread agreement among statisticians.
However, in the next subsection, we show that the set $\{R,P_R,F_R\}$ is a sufficient statistic for all generative statistical models with constant image $T$.
We do this by showing that any likelihood evaluation of every constant image $T$ can be done using only $\{R,P_R,F_R\}$.
Therefore any decision between any two statistical models, even composite models, can be done using $\{R,P_R,F_R\}$ alone.
%
%
%

\subsection{$\{R,P_R, F_R\}$ are sufficient for any measurement}
\label{subsec:Measurement}
Assuming we want to measure some parameters $\theta$ of a generative model of the true image $T$ (e.g., flux measurement of a star or fitting a galaxy model).
Having a generative model means that each value of $\theta$ predicts some $T_\theta$. 
The measurement process, whether Bayesian or frequentist, relies only on the calculation of the log-likelihood function $\mathcal{L}(\theta)$ for all values of $\theta$.
In order to show that $\{R,P_R,F_R\}$ are sufficient, we will first show that the exact numerical value (of the part that depends on $\theta$) of any likelihood evaluation of the data, given some true image $T_\theta$ can be computed by using only $\{R,P_R,F_R\}$.

Following this, we prove that $\{R,P_R,F_R\}$ are sufficient for any measurement of a constant in time property, using the Fisher-Neyman factorization theorem \citep{Fisher,Neyman}. 
This theorem states that a function $\mathcal{F}$ of the data $M$, is sufficient for $\theta$ if and only if it satisfies that the log-likelihood $\mathcal{L}$ of the data given $\theta$ factorizes into:
\begin{equation}
\mathcal{L}(M|\theta) = f(M) + g_\theta(\mathcal{F}(M))\,,
\end{equation}
where $f$ is an arbitrary function that its argument is independent from the model parameters and $g_\theta$ is an arbitrary function that depends on the data only thorough the statistic $\mathcal{F}$.
We are going to show this for any generative model parametrized by $\theta$ for the constant in time true image $T(\theta)$, and by that, show that $\{R,P_R,F_R\}$ are sufficient for {\it any} time invariant measurement on the data.

The log of the probability to measure the data given a value of $\theta$ for the generative model is (using the assumption of additive white Gaussian noise): 
\begin{align}
\mathcal{L}(M|\theta) =& \sum_j{\ln(\mathcal{P}(M_j|\theta))} \\ =& \sum_j{\sum_x\frac{\left|M_j(x) - (F_jT_\theta\otimes P_j)(x)\right|^2}{2\sigma_j^2}}\,. \end{align}
Using the Parseval equality, we can express the likelihood
as a function of $\theta$ by:
\begin{align} \mathcal{L}(M|\theta) = \sum_j{\sum_f\frac{\left|\widehat{M_j}(f) - (F_j\widehat{T_\theta} \widehat{P_j})(f)\right|^2}{{2\sigma_j^2}}}.
\end{align}
Opening the squares, and keeping only terms that depend on $\theta$ we get:

\begin{align}
\mathcal{L}(\theta)&= \sum_f|\widehat{T_\theta}(f)|^2\left(\sum_j\frac{|F_j\widehat{P_j}(f)|^2}{2\sigma_j^2}\right) -   2\Re\left[\sum_f\overline{\widehat{T_\theta}}\sum_j{\widehat{M_j}\frac{F_j\overline{\widehat{P_j}}}{2\sigma_j^2}}\right] \\ \nonumber 
& = \frac{1}{2}\sum_f{\left(|\widehat{T_\theta}|^2F_R^2|\widehat{P_R}|^2 \right)(f)} - \Re\left[\sum_f{\left(\overline{\widehat{T}_\theta} \widehat{S}\right)(f)}\right]\,,
\end{align}
where $\mathcal{R}$ denotes the real part operator.
Furthermore, by adding the constant term $\sum_f |\widehat{R}|^2$, that does not depend on $\theta$, we can further simplify this to (using Eq. \ref{eq:RelationSR}):
\begin{align}
\mathcal{L}(M|\theta)+ f(M) &= \sum_f \left(|\widehat{R}|^2 +|\widehat{T_\theta}|^2F_R^2|\widehat{P_R}|^2  - 2\Re\left[\overline{\widehat{T}_\theta F_R\widehat{P_R}} \widehat{R}\right]\right) \\ \nonumber &= \sum_f{\left|\widehat{R} - F_R\widehat{P_R}\widehat{T_\theta} \right|^2}\,. \end{align} Using the Parseval equality again, we get:
\begin{align}\label{eq:Suff} \mathcal{L}(M|\theta) = \sum_x{\frac{\left(R - F_RP_R\otimes T_\theta \right)^2}{\sigma_R^2}}\,.
\end{align}
The fact that the expression we derived depends on the data only through $\{R,P_R,F_R\}$ proves, using the Fisher-Neyman factorization theorem that they are indeed sufficient for measuring the parameters of any generative model.

From examining Equation \ref{eq:Suff}, we can see another important property of $\{R,P_R,F_R\}$, which is that the measurement process using them is {\it exactly} the same as if $R$ was a regular single image, $P_R$ its PSF and $F_R$ is its transparency.
This means that performing measurements on $\{R,P_R,F_R\}$ is indistinguishable from performing measurements on regular observations. Therefore any existing code that receives a single image could be used unchanged on $R$, getting $P_R$ as the relevant PSF.
This is the main reason why we prefer the set $\{R,P_R,F_R\}$ over $\{S,P_R,F_R\}$ even though they are statistically equivalent. 
There is another reason, which is the same reason why astronomers usually use images, and not matched-filtered images -- they are more intuitive to interpret.

\section{Examples}\label{sec:ExamplesVisualization}
In \S\ref{sec:Sufficiency} we presented major advantages of proper image coaddition.
These are optimal $S/N$, and the ability to perform any measurement or decision optimally without the original images.
Another advantage of this method, as we mentioned earlier,
is that it has a PSF which is typically sharper
than the sharpest image in the original set.
In order to demonstrate these advantages of the new
method over the popular methods we present here a few
simulated examples (\S \ref{subsec:Visualization}) as well as coaddition of real seeing-limited data (\S\ref{subsec:realData}) and speckle images (\S\ref{subsec:realDataSpeckles}).
Additional examples, more specific to fast imaging applications, with a much more quantitative analysis,
will be presented in future papers of this series. 
\subsection{Visualization of the $R,P_R$ statistics for various special cases} \label{subsec:Visualization}
\begin{figure*}

\centering
\includegraphics[width=120mm]{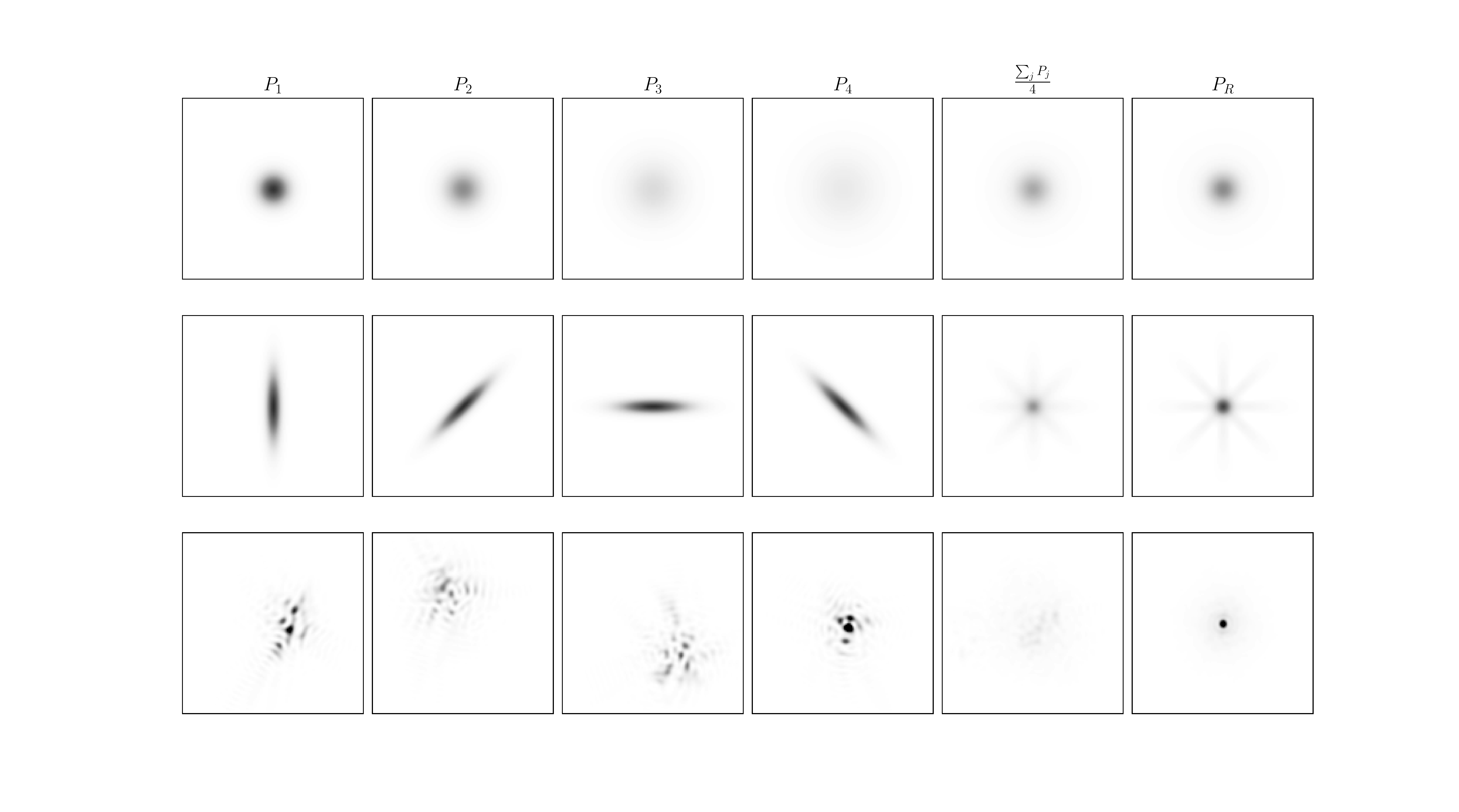}
\caption{An image illustrating the point spread functions (PSF) of properly added images.
The four left columns: Individual simulated PSFs.
Second from the right: Regular equal weights coaddition.
Right: The PSF of the proper coaddition image (i.e., using Eq. \ref{eq:Prhat}).
On top: Gaussian PSFs with widths of $8$, $10$, $15$ and $18$ pixels respectively.
Middle: PSFs are elliptical Gaussians with various orientations and a width of $5$ pixels and $25$ pixels on the short and long axes, respectively.
Bottom: simulated speckle PSFs, corresponding to $\frac{D}{r_0}=10$, where $D$ is the telescope diameter and $r_0$ is the Fried length. In the case of speckle images, $20$ images were coadded, but only 4 of the original images are displayed. Coadded images on the same row are displayed with the same color map. \label{fig:PSFs}}
\end{figure*}

\begin{figure*}

\centering
\includegraphics[width=120mm]{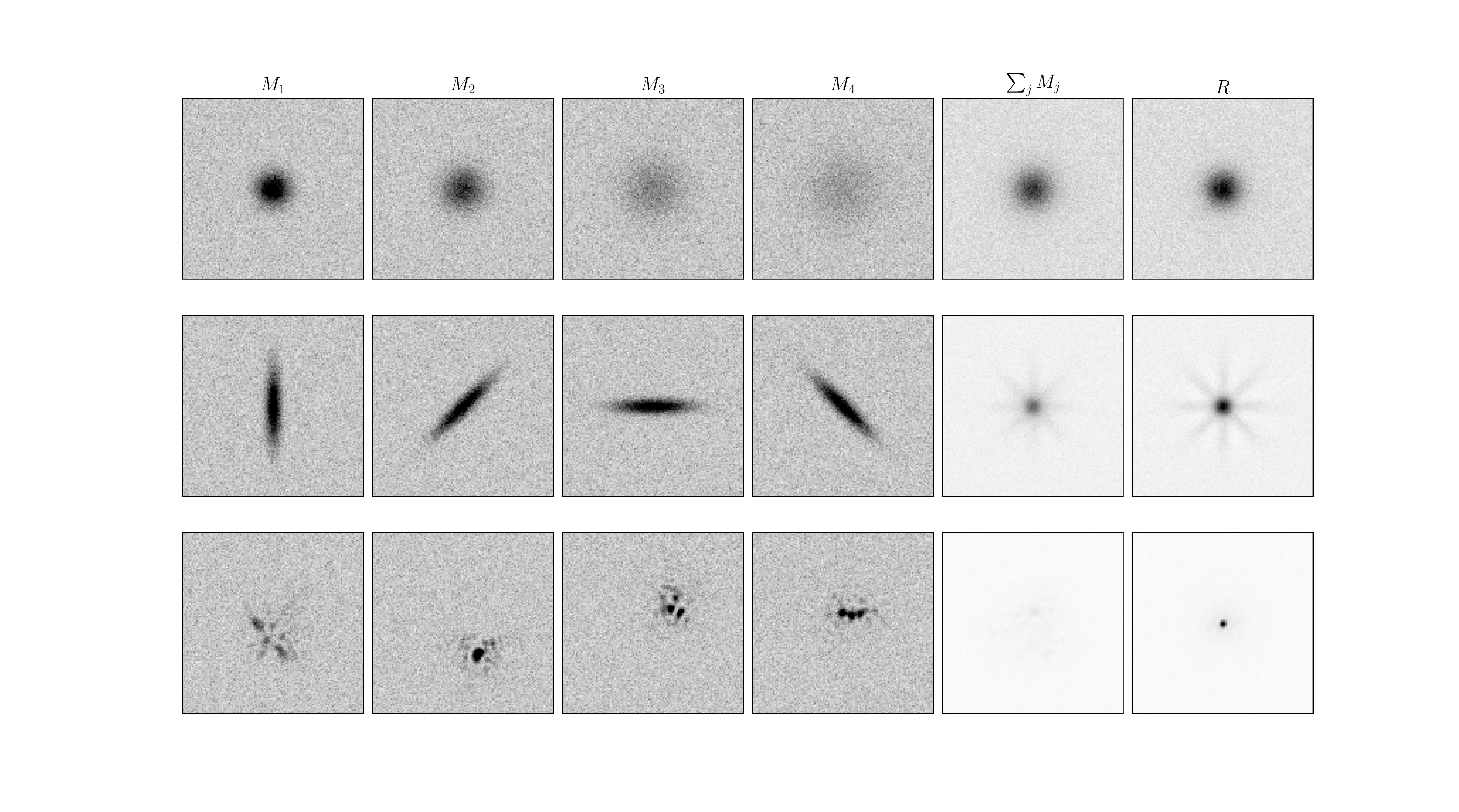}
\caption{Like Fig.~\ref{fig:PSFs}, but showing the images, with noise, rather than the PSFs (i.e., using Eq. \ref{eq:Rhat}). Coadd images are shown in the same gray scale, and are normalized to have the same noise standard deviation. \label{fig:images}}
\end{figure*}

\begin{figure*}

\centering
\includegraphics[width=120mm]{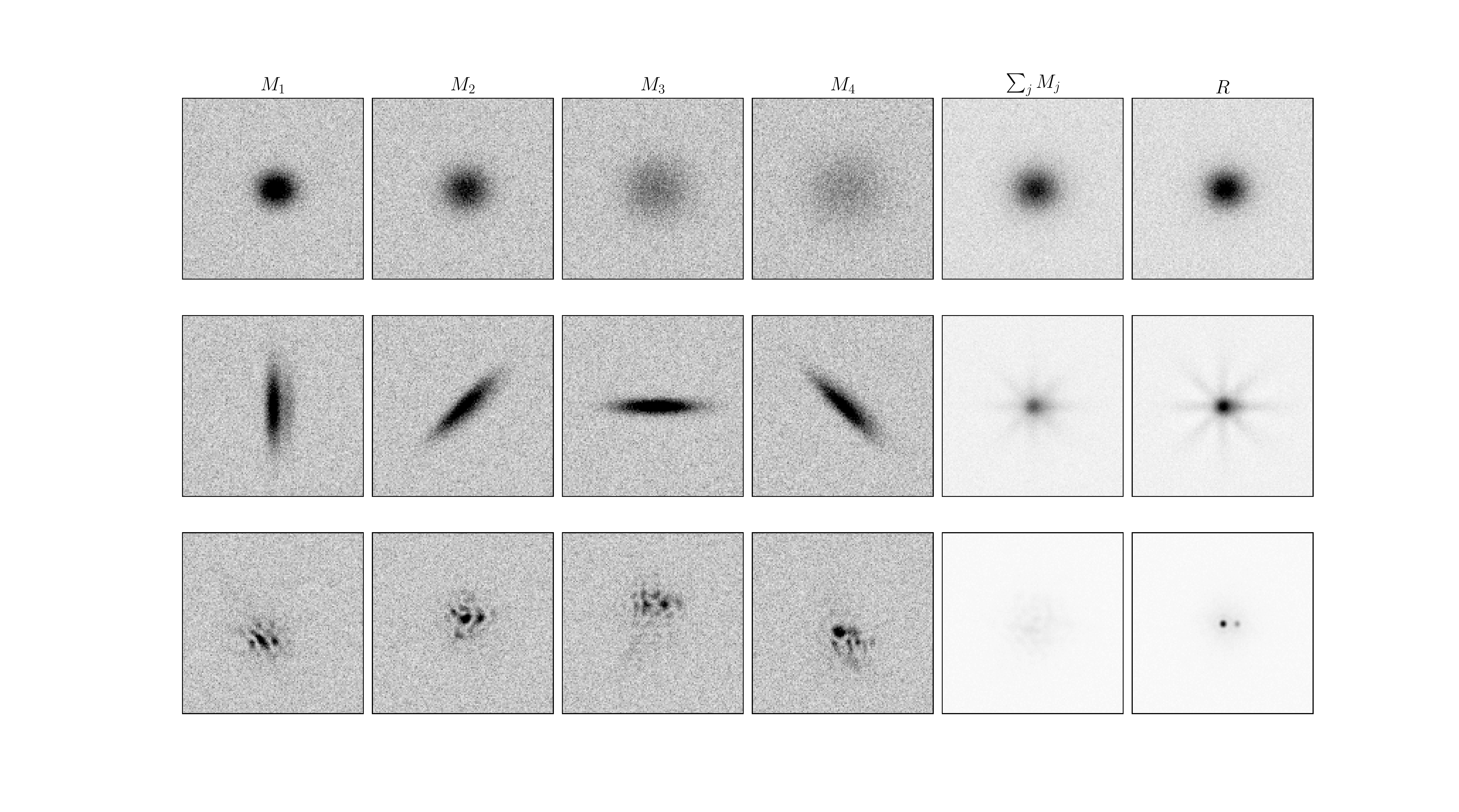}
\caption{Same as Fig.~\ref{fig:images}, but where the source is a binary star with separation of 10 pixels and flux ratio of 2.5. \label{fig:binarys}}
\end{figure*}

Here we show visual examples of the results of the coaddition process, performed on simulated data.
We directly compare these results to the naive direct sum of images.
In order to differentiate the effect of the PSF shape from that of
the transparency and background, we use equal transparency ($F_{j}$)
and background variance ($B_{j}$) for all images coadded.
We note that under these conditions even the more elaborate weighted addition methods \citep{Annis2014,Jiang2014} boil down to a direct sum.
%
Figure~\ref{fig:PSFs} compares the PSFs of the direct sum with that of the proper image (Eq. \ref{eq:Prhat}), while Figure \ref{fig:images} shows the resulting differences in the images of a point source, with noise, between the direct sum and proper coaddition (Eq. \ref{eq:Rhat}).
Figure~\ref{fig:binarys} shows the same as Figure~\ref{fig:images} but for a binary star with a separation of 10 pixels and 1 magnitude flux difference. 
Each row in Figure~\ref{fig:PSFs} shows, left to right,
four PSFs (denoted $P_{1}$ to $P_{4}$), the PSF resulting from simple
coaddition $\left(\frac{\sum_j{P_{j}}}{N_{\rm im}}\right)$, where $N_{\rm im}$ is the number of images coadded, and the PSF of the proper image ($P_R$; right column).
The first row corresponds to Gaussian PSFs with varying widths (i.e., varying standard deviations),
where the Gaussian standard deviations are 8, 10, 15 and 18 pixels, respectively.
The second row shows highly elliptical PSFs with varying orientations, while the third row shows speckle images of a point source, simulated with
$D/r_{0}=10$, where $D$ is the telescope diameter and $r_{0}$ is the Fried length \citep{Fried66}.
Apart from the visually better PSF, properly coadding images results in lower noise standard deviation relative to other methods. This effect can be seen in Figures \ref{fig:images} and \ref{fig:binarys}. In these figures, we compare naive coaddition ($\sum_j{M_j}$) and the proper image ($R$) of noisy simulated images with PSFs chosen as in Figure~\ref{fig:PSFs}, both normalized to have a noise standard deviation of 1 and are shown with the same gray scale.
It is clearly seen that both the shape and the signal to noise of the images are better in the properly coadded image than in the naive summation of the images.

\begin{figure*}

\centering
\includegraphics[width=170mm]{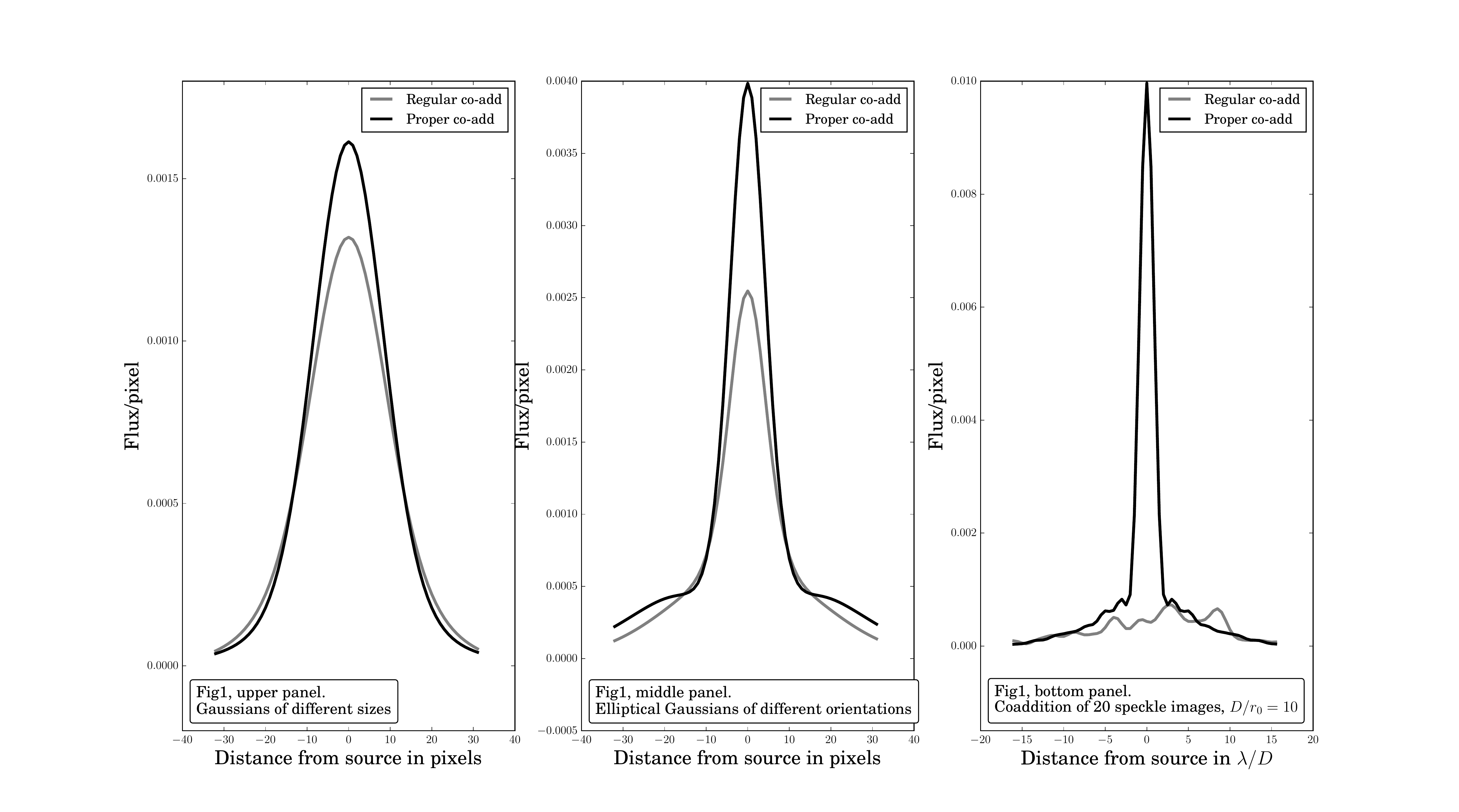}
\caption{Intensity values of the point spread functions presented in Figure \ref{fig:PSFs} along a line crossing at its center.
\newline
Black -- The PSF of the properly added images.
Grey: -- The PSF of the regularly summed images.
\newline
Left: cross cuts for the top row (circular Gaussians of different widths).
\newline
Middle: cross cuts for the middle row (elliptical Gaussian PSFs with various orientations).
Right: cross cuts for the bottom row (speckle images).
\label{fig:PSFsprofile}}

\end{figure*}

\begin{figure*}

\centering
\includegraphics[width=170mm]{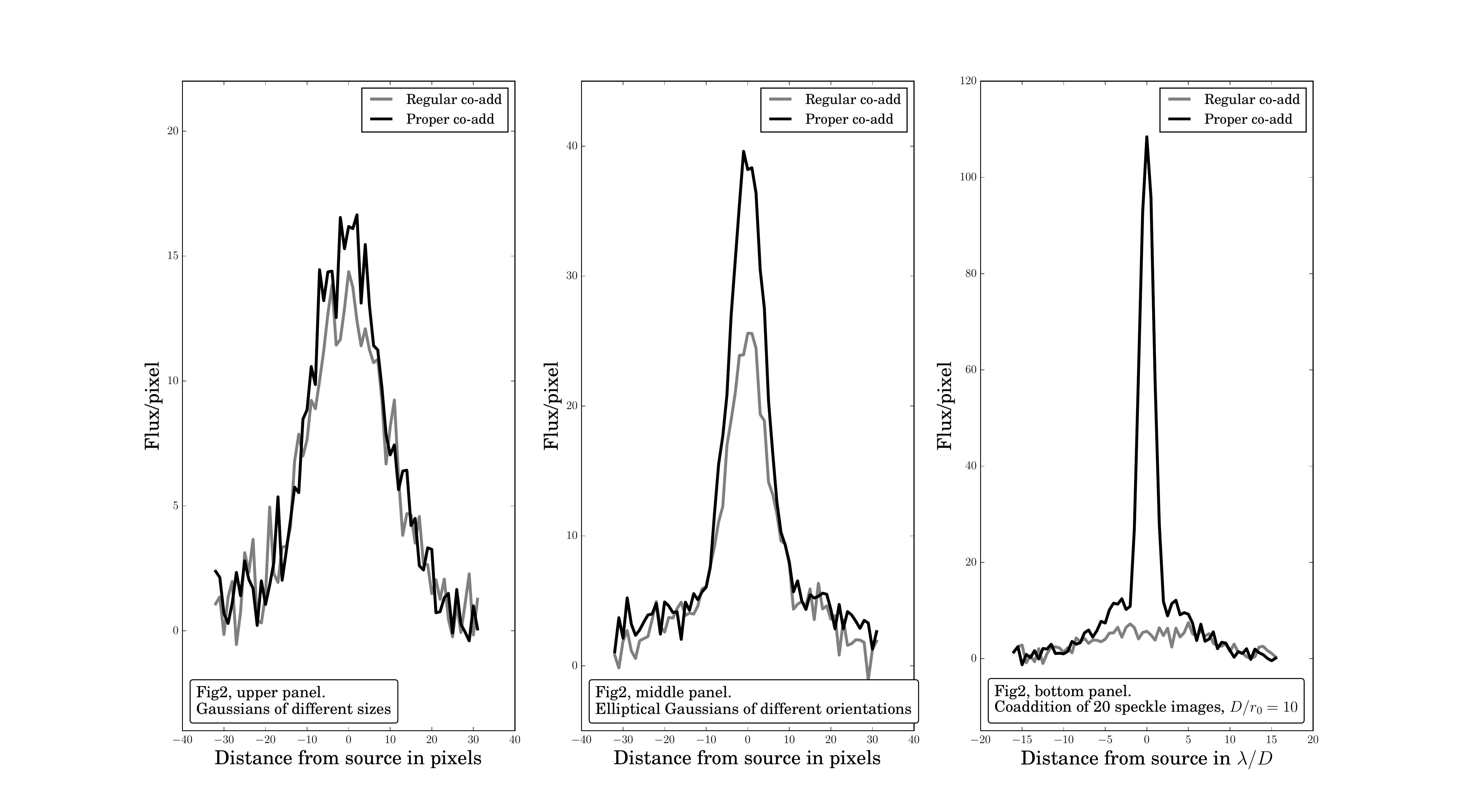}
\caption{Like Fig. \ref{fig:PSFsprofile}, but corresponding to the images presented in Figure \ref{fig:images}.
\label{fig:Imagesprofile}}

\end{figure*}

\begin{figure*}

\centering
\includegraphics[width=170mm]{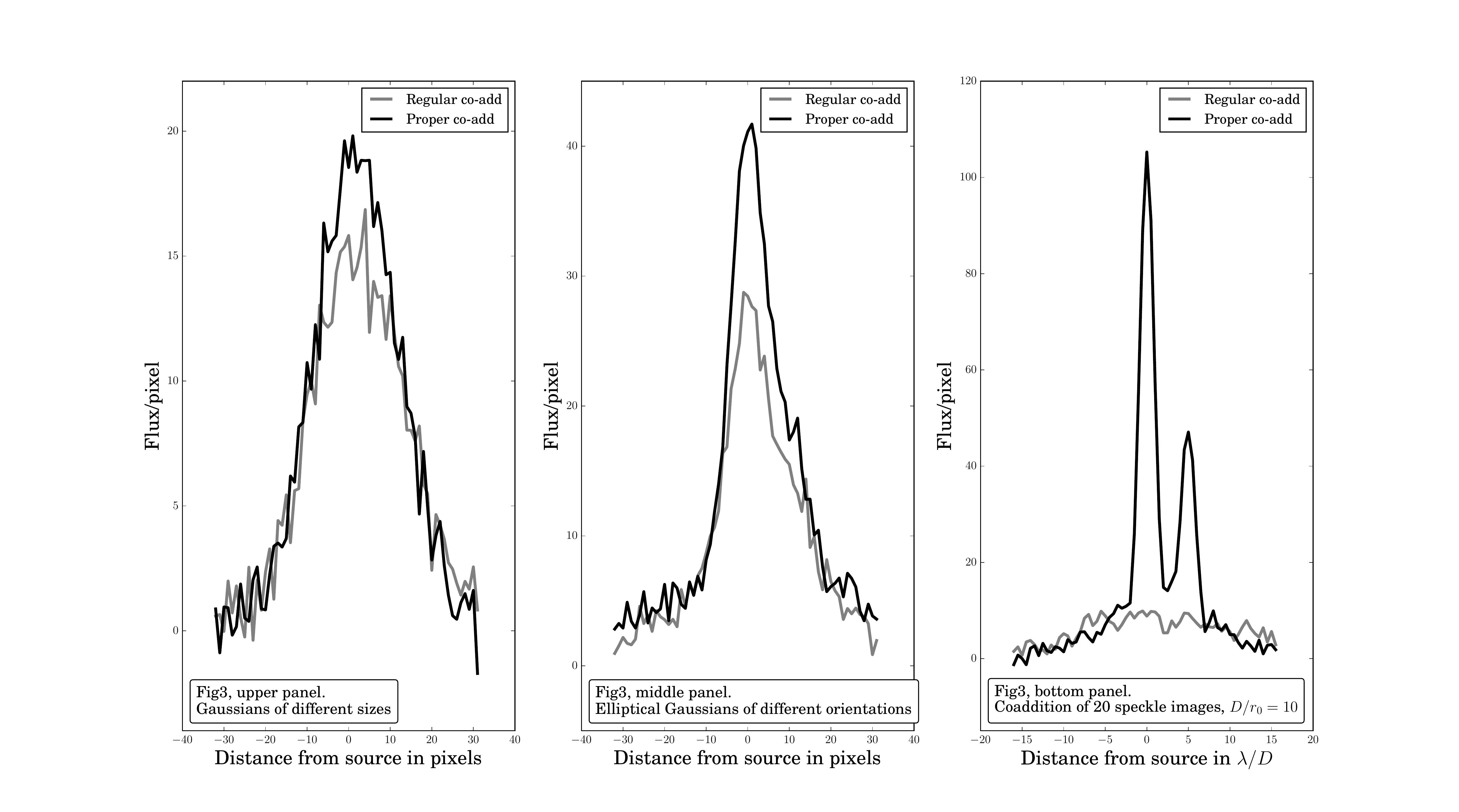}
\caption{Like Fig. \ref{fig:PSFsprofile}, but corresponding to the images presented in Figure \ref{fig:binarys}.
\label{fig:Binarysprofile}}

\end{figure*}

To aid comparison, we plot the central horizontal cuts through all coadd images presented in Figures~\ref{fig:PSFs}, \ref{fig:images} and \ref{fig:binarys} in Figures ~\ref{fig:PSFsprofile}, \ref{fig:Imagesprofile} and \ref{fig:Binarysprofile}, respectively.
The gray lines show the result of simple coaddition, while the black lines show the result of proper coaddition.

We note that our coaddition technique is optimal by construction, and is not assuming anything on the images themselves (except having background dominated, uncorrelated Gaussian noise). Therefore, it works optimally on any complex image. An example for this is given in Figure~\ref{fig:Binarysprofile}, in which the input image was a simulated binary star. It is very clear that the 1 magnitude fainter companion can be easily detected by further signal processing of the proper image in both the asymmetric PSF case (second row) and speckle PSF case (third row).

The coaddition method we present will provide the largest improvement in comparison to weighted coaddition when there is a large diversity in the PSF shape between the coadded images.
In our examples, the most prominent effect is seen in the case of the speckle images.
It is further striking, that the proper coaddition of speckle images generates an image containing substantial spatial information right to the diffraction limit (see demonstration on real images in \S \ref{sec:ExamplesVisualization}). Given the importance of this subject we discuss this in detail, under different scientifically relevant observational situations, in future papers.

\subsection{Tests on real seeing-limited observations}
\label{subsec:realData}

For source detection, the method we present in this paper is identical
to the technique outlined in paper~I.
The reason for this is that in order to find sources in our optimal coadd image (Eq.~\ref{eq:Rhat}),
we need to cross correlate the image with its PSF (Eq.~\ref{eq:Prhat}).
Doing this will result in Equation~\ref{Eq.S}
which is simply the method we presented in paper~I.
Nevertheless, in practice numerical errors, and sub-optimal estimation
of the various properties (e.g., PSF) may slightly change the outcome.
Therefore, it is important to demonstrate that in practice the method presented here
returns the same results as the optimal coaddition for source detection in paper~I,
and that these results are better than those obtained using other techniques.

As in paper~I, we test our method on data
from the Palomar Transient Factory (PTF; \citealp{PTF,PTF2}) Data Release 2.
The image reduction process is described in \cite{Laher14},
and photometric calibration is discussed in \cite{Ofek12}.
Furthermore, we use the same datasets and tools as before.
We used four datasets, and for each dataset we constructed a deep reference image,
from over 400 good seeing images, coadded using the \cite{Annis2014} method.
Next, for each dataset we choose small subsets, containing 19 to 45 random-seeing images,
and coadded these images using various methods including
equal weights, \cite{Annis2014} weighted coaddition,
\cite{Jiang2014} coaddition and our proper coaddition technique.
The selection criteria for each data set are listed in Table~\ref{tab:Selection}.
For more details we refer the reader to paper~I.


\begin{deluxetable*}{llllll}
\tablecolumns{5}
\tablewidth{0pt}
\tablecaption{Sets of images used in the comparison}
\tablehead{
\colhead{Set}         &
\colhead{Field}       &
\colhead{CCD}         &
\colhead{\#im}         &
\colhead{Type}        &
\colhead{Criteria}     \\
\colhead{}            &
\colhead{}            &
\colhead{}            &
\colhead{}            &
\colhead{}            &
\colhead{}
}
\startdata
1 &  100031 & 6 & 425  & Ref.   & Variance$<$1000\,e$^{-}$ \& FWHM$<$4'' \& $>10$ PSF stars \\
  &         &   &  45  & coadd  & All images taken on Oct, Nov, Dec 2012 \& $>10$ PSF stars \\
\hline
2 &  100031 & 6 & 425  & Ref.   & Variance$<$1000\,e$^{-}$ \& FWHM$<$4'' \& $>10$ PSF stars \\
  &         &   &  48  & coadd  & All images taken on the first 9 days of each Month in 2011 \& $>10$ PSF stars \\
\hline
3 &  100031 & 4 & 263  & Ref.   & Variance$<$1000\,e$^{-}$ \& FWHM$<$4'' \& $>10$ PSF stars \\
  &         &   &  39  & coadd  & All images taken on Oct, Nov, Dec 2012 \& $>10$ PSF stars \\
\hline
4 &  100031 & 4 & 263  & Ref.   & Variance$<$1000\,e$^{-}$ \& FWHM$<$4'' \& $>10$ PSF stars \\
  &         &   &  17  & coadd  & All images taken on the first 9 days of each Month in 2011 \& $>10$ PSF stars 
\enddata
\tablecomments{Selection criteria for reference images and coadd images
in the various sets used for testing the coaddition methods.
Field indicate the PTF field ID,
while CCDID is the CCD number in the mosaic (see \citealp{PTF,Laher14}).
Type is either Ref. for the deep reference image, or Coadd for the subset
we coadd using the various methods.
Note that field 100031 is in the vicinity of M51.
This table is identical to Table~1 in paper~I.
\label{tab:Selection}}
\end{deluxetable*}

We note that there are two ways to calculate the PSF of the proper coaddition image.
The first is to measure it from the proper coaddition image,
while the second is to calculate it using Equation~\ref{eq:Rhat},
and the PSFs measured in the individual images.
Using the test data presented in paper~I, we find that the second method
gives much better results than the first method. This is likely due to the way that errors are
propagated in the two schemes, and means that storing $P_R$ is recommended.

We ran several tests on the various coadd images.
We ran our source extraction code (see paper~I for details) on the various images,
and matched the sources against those found in the deep reference image.
This allowed us to quantify how many real stars and how many false detections
are reported based on each coaddition method.
The number of real sources and false sources found in each image in each dataset
is given in Table~\ref{tab:Comp}.

\begin{deluxetable}{lllllll}
\tablecolumns{7}
\tablewidth{0pt}
\tablecaption{Comparison between coaddition methods}
\tablehead{
\colhead{Set}               &
\colhead{Method}            &
\colhead{$I_{{\rm r}}$}            &
\colhead{$N_{{\rm d}}$}     &
\colhead{$N_{{\rm r}}$}     &
\colhead{$N_{{\rm u}}$}     &
\colhead{$N_{{\rm f}}$}\\
\colhead{}            &
\colhead{}            &
\colhead{}            &
\colhead{}            &
\colhead{}            &
\colhead{}            &
\colhead{}
}
\startdata
1 & Deep            &      0  &       2433    &   \nodata  &    \nodata   &   \nodata \\
  & Equal weights   & 0.65    &       1136    &     1087   &      1346    &       49  \\
  & \cite{Annis2014}&0.91     &       1332    &     1255   &      1178    &       77  \\
  & \cite{Jiang2014}&0.89     &       1299    &     1229   &      1204    &       70  \\
  & paper~I       &      1  &       1417    &     1324   &      1109    &       93  \\
  & This work        &      1  &       1419    &     1325   &      1108    &       94  \\
\hline
2 & Deep            &      0  &       2433    &   \nodata  &    \nodata   &   \nodata \\
  & Equal weights   &0.59     &       1183    &     1085   &      1348    &       98  \\
  & \cite{Annis2014}&0.93     &       1481    &     1343   &      1090    &      138  \\
  & \cite{Jiang2014}&0.90     &       1421    &     1301   &      1132    &      120  \\
  & paper~I       &      1  &       1541    &     1390   &      1043    &      151  \\
  & This work        &      1  &       1542    &     1391   &      1042    &      151  \\
\hline
3 & Deep            &      0  &       4062    &   \nodata  &    \nodata   &   \nodata \\
  & Equal weights   &0.53     &       1645    &     1539   &      2523    &      106  \\
  & \cite{Annis2014}&0.84     &       1955    &     1796   &      2266    &      159  \\
  & \cite{Jiang2014}&0.81     &       1912    &     1759   &      2303    &      153  \\
  & paper~I       &      1  &       2206    &     1976   &      2086    &      230  \\
  & This work        &      1  &       2206    &     1976   &      2086    &      230  \\
\hline
4 & Deep            &      0  &       4062    &   \nodata  &    \nodata   &   \nodata \\
  & Equal weights   &0.83     &       2319    &     2007   &      2055    &      312  \\
  & \cite{Annis2014}&0.82     &       2144    &     1981   &      2081    &      163  \\
  & \cite{Jiang2014}&0.92     &       2272    &     2062   &      2000    &      210  \\
  & paper~I       &      1  &       2401    &     2125   &      1937    &      276  \\
  & This work        &0.99     &       2400    &     2124   &      1938    &      276  
\enddata
\tablecomments{Quantitive comparison between images coadded using various technique
matched against a deep reference image.
The four blocks, separated by horizontal lines, correspond to the four data sets in Table~\ref{tab:Selection}.
$I_{{\rm r}}$ is the approximate survey speed ratio between the image compared and our optimal coaddition method (see text for details).
$N_{{\rm d}}$ is the number of detected sources to detection threshold of 4$\sigma$.
$N_{{\rm r}}$ is the number of real sources (i.e., detected sources which have a counterpart
within $3''$ in the deep reference image.
$N_{{\rm u}}$ is the number of real undetected sources (i.e., sources detected in the reference which are not detected in the coadd image).
$N_{{\rm f}} = N_{d}-N_{r}$ is the number of false detections.
This table is identical to Table~2 in paper~I.
\label{tab:Comp}}
\end{deluxetable}

Furthermore, like in paper~I, we compared the detection $(S/N)^{2}$ (proportional to the survey speed
and information content) of the sources in the various images.
For each star detected in the reference image we divide its $S/N$ measured
in a coaddition image by its $S/N$ measured in our properly coadded image.
Next we take the median of the squares of these ratios.
This roughly gives the survey speed of each coadded image,
relative to our coaddition method, and is listed in Table~\ref{tab:Comp} ($I_{r}$).

Table~\ref{tab:Comp} indicates that our proper coaddition method returns the same results, up to some small numerical differences,
as the optimal coaddition for source detection presented in paper~I.
Next it is clear that our method perform better than the other popular methods.

\begin{figure}
\centering
\includegraphics[width=80mm]{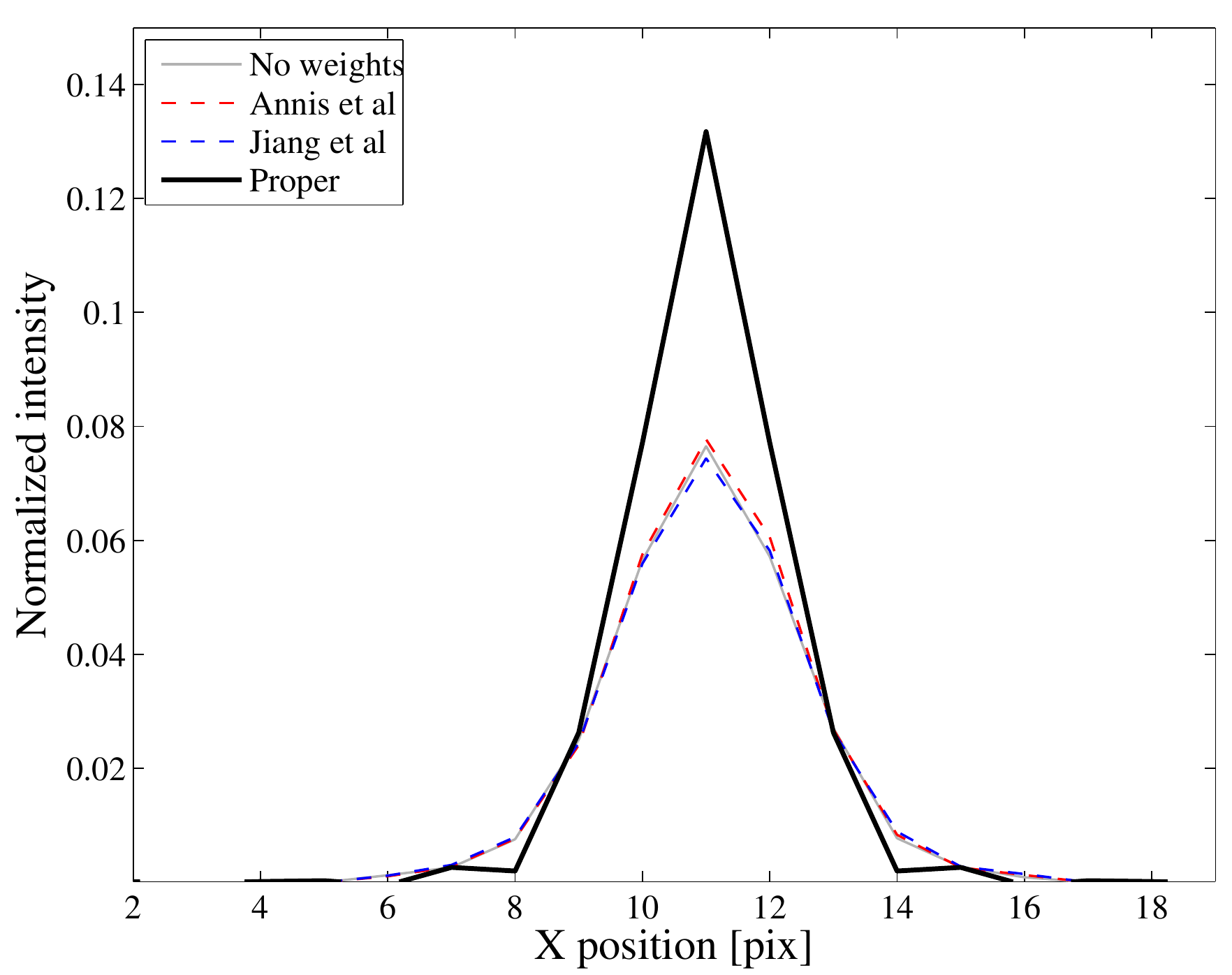}
\caption{A cut through the PSF of the various methods
based on dataset 3 from paper~I.
Shown are the proper coaddition image (black line),
the unweighted addition (gray line),
\cite{Annis2014} weighted addition (dashed red line),
and \cite{Jiang2014} weighted addition (dashed blue line).
All the PSFs are normalized to have unity sum.
The PSF of the proper coaddition was calculated using Equation~\ref{eq:Prhat}.
\label{fig:realPSF}}
\end{figure}

Another important point is that the effective PSF of our method is typically
sharper than the PSF of other methods.
Figure~\ref{fig:realPSF} shows an example of a cut through the PSFs of various
image coaddition techniques (using dataset 3).
The proper image coaddition (black line)
has narrower and higher PSF than other methods.
This demonstrates, as shown in \S\ref{sec:BestProperImage} and \S\ref{sec:Sufficiency},
that the proper image indeed contains more spatial information than existing methods,
and provides significant improvement to the quality of all further signal processing
of the coadd image, both in theory and in practice.
To summarize, the method presented in this paper, provide a few percents to 25\%
improvement in terms of survey speed (or information content)
over popular weighted coaddition techniques (e.g., Annis et al. 2014; Jiang et al. 2014; see paper~I for details).
Furthermore, the effective PSF of the coadded images is typically sharper than
those provided by popular image coaddition methods.

\subsection{Tests on real speckle images}
\label{subsec:realDataSpeckles}

As evident from the third row of Figures~\ref{fig:PSFs},\ref{fig:images} and \ref{fig:binarys}, proper image coaddition
has a great potential for high resolution imaging.
Unlike lucky imaging \citep{Law06} this method uses the best possible linear combination
of {\it all} the images.
Furthermore, unlike speckle interferometry \citep{SpeckleInterferometry}
and deconvolution from wave-front sensing (DWFS; \citealp{DWFS}),
this method is numerically stable and
does~not invent non-existing information, or introduce long range correlations into the data.
It is worth while to explain these two points.
In the case of speckle interferometry, due to the division by small numbers in the Fourier space, the noise
component in poorly known spatial frequencies is amplified, and becomes correlated noise in the real space.
This noise limits the contrast of speckle interferometry as well as any
honest significance analysis of detected companions.
This is a major reason why astronomers resort to visual inspection of the auto-correlation
maps produced by speckle interferometry for binary detection.
Furthermore, because deconvolution methods enforce a delta function PSF
for the images, they invent non-existing information about poorly known spatial frequencies (e.g., above the Nyquist frequency).

In future papers we will discuss our new de-speckling techniques.
These techniques are based, with some modifications for source dominated noise and unknown PSF,
on Equation~\ref{eq:Rhat}.

The formula we derived in this paper is optimal only for the background-noise-dominated case,
but it is worth while to visualize its performance as is for real images,
even if they are source-noise dominated.
Here we wish to briefly demonstrate that this method has an excellent potential for
high contrast imaging via speckle observations, but for additional details,
as well as new mathematical derivations, improvements and comparisons to existing methods, see future papers.

\begin{figure*}
\centering 
\includegraphics[width = 160mm]{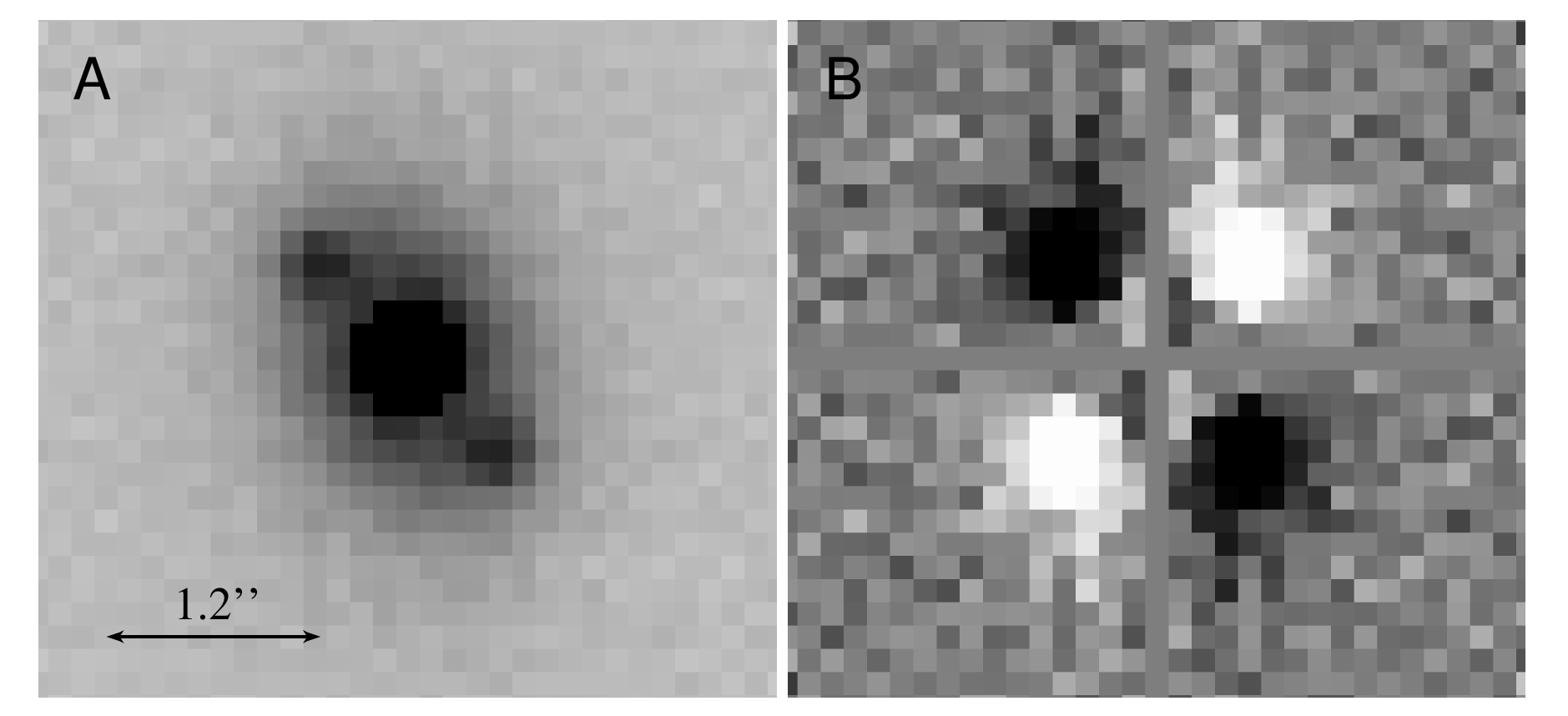}
\caption{{\bf Panel A}: A proper coaddition of 30,000 7\,ms exposures (speckle images)
of the binary star COU1049.
The images were taken under seeing of 1.2''-1.5'' FWHM.
The companion, at angular distance of $0.75''\pm0.1''$ is clearly detected,
along with its mirror image.
The mirror image is generated due to the fact that we used the image
as its own PSF.
The FWHM of the resulting image is about 0.35''
(note that it is comparable to the diffraction limit; $\lambda/D\cong0.3''$).
{\bf Panel B}: A symmetric subtraction of the coadded image with respect to the central vertical axis. The PSF is expected to be symmetric, therefore this removes the primary star's contribution. We are left with two positive and two negative copies of the companion.
}
\label{fig:speck}
\end{figure*}

\begin{figure}
\centering 
\includegraphics[width = 80mm]{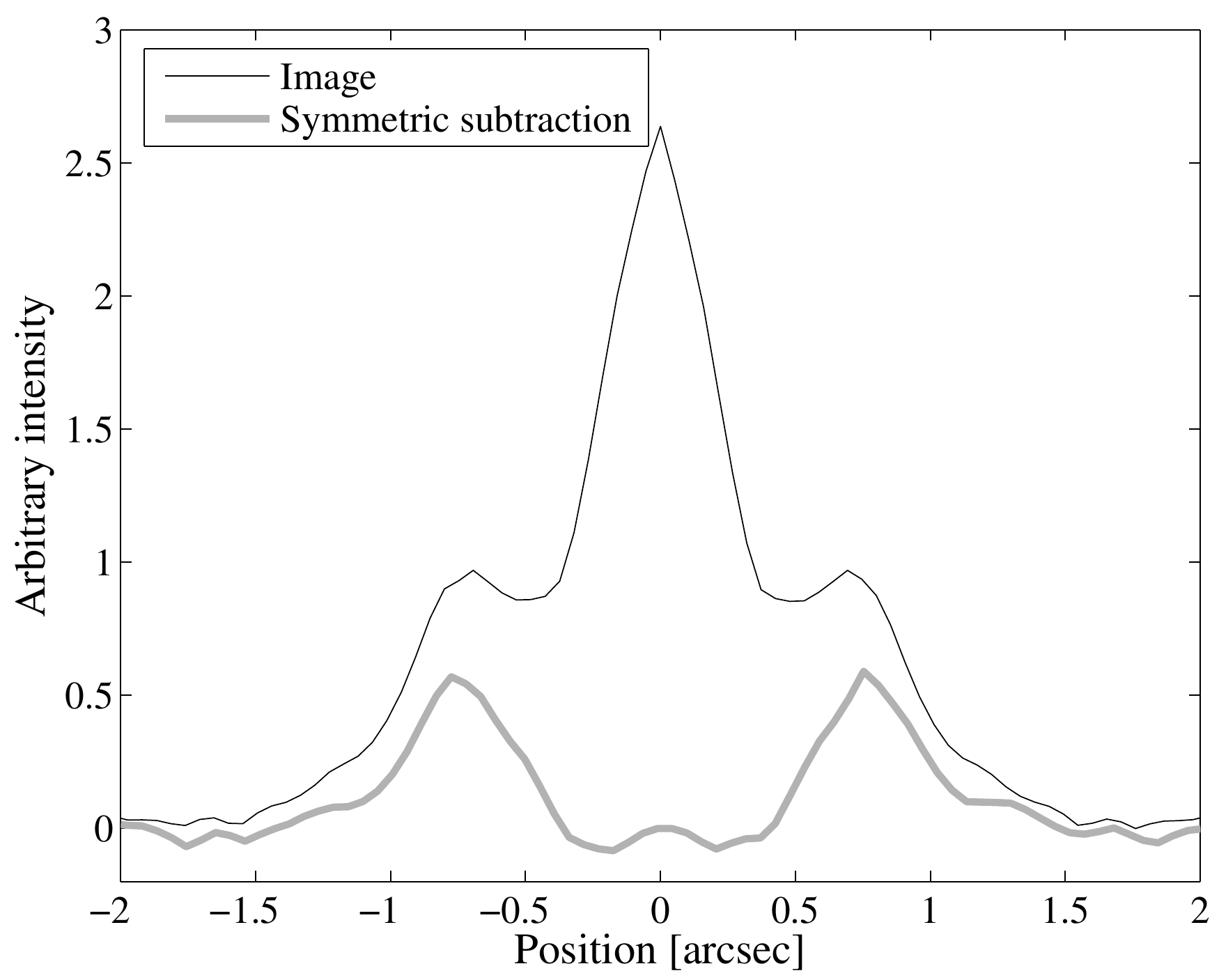}
\caption{A cut through the COU1049 star and its companion shown in Figure~\ref{fig:speck}. Black line is for Panel A while the gray line is for Panel B.
The achieved seeing FWHM is 0.35'' (see Fig.~\ref{fig:speck} for details),
which is substantially smaller than the 1.2''-1.5'' FWHM in the original images.}
\label{fig:speck_prof}
\end{figure}

To test this we obtained 30000 7\,ms images of the star COU 1049 (RA $=00^{h}29^{m}45^{s}.34$, Dec $=+36^{\circ}50^{'}23^{''}.7$).
The images were taken using
the Kraar Observatory\footnote{http://www.weizmann.ac.il/particle/kraar/home} 40cm telescope equipped with
an Andor/Zyla sCMOS camera.
The camera has 6.5\,$\mu$m pixels, and we observed using a Barlaw lens, extending the focal ratio to f/25, so the pixel scale was 0.13''\,pix$^{-1}$.
This pixel scale is smaller than the Nyquist frequency of the optics, allowing for diffraction limited resolution
if the best possible correction is applied. 
We note that this is not an essential requirement for the application of the method.
The images were taken under seeing conditions of 1.2''-1.5'' full width half max.
This is equivalent to $D/r_{0}\approx 5$,
where $D$ is the telescope aperture and $r_{0}$ is the Fried length \citep{Fried66}.
Figure~\ref{fig:speck} presents the coadd image using Equation~\ref{eq:Rhat}.
As the PSF, we used the image itself.
COU1049 is a binary star system, with separation of 0.7''.
Its primary and secondary magnitudes are 9.8 and 10.15, respectively.
The de-speckled image created by proper coaddition
clearly shows the companion, along with its mirror image.
This mirror image is generated due to the fact we used the image as its own PSF.
Figure~\ref{fig:speck_prof} shows a cut through the PSFs of the two components.
First it is evident that the proper coaddition of speckle images
reveals diffraction limited information.
Quantitative and qualitative comparison of this method to existing speckle imaging methods are left to future papers.

%
\section{Code and implementation details}
\label{sec:Code}

The code we use to implement the proper image coaddition is similar to the code presented in paper~I.
The only additional program required is the MATLAB function {\tt sim\_coadd\_proper.m}.
This function gets the images (optionaly not aligned),
their weights and PSFs, and returns the proper coadded image and its PSF,
where the PSF is calculated using Equation~\ref{eq:Prhat},
rather than measuring it from the proper coadd image.
All the calibration steps (registration and PSF, background, and variance estimation)
were done using the same programs described in paper~I.
All the code is available as part of the MATLAB astronomy and astrophysics
package\footnote{http://webhome.weizmann.ac.il/home/eofek/matlab/} (Ofek 2014).

We note that the usual implementation problems like
registering the images, resampling them to the same grid,
measuring the background and variance levels, and measuring/interpolating
the PSF, as always, require attention.
However, the attention to the details given this method is not different than in the case of other methods.
Nevertheless, we suggest that care will be given to the following points:

{\bf Taking care of defects in the images:} 
It is important to note that this method is not generalized yet to
use robust estimators like median or sigma clipping (see discussion in paper~I).
Therefore, one needs to take care of image artifacts (bad pixels, cosmic rays, ghosts)
prior to the coaddition process.
We recommend on detecting these artifacts prior to the coaddition, and interpolating
over the bad pixels.
This can be done using existing algorithms like \cite{Dokkum,Ipatov,Ritter}.
Alternatively, cosmic ray removal can be done using image subtraction.
However, this requires a sensitive image subtraction method which is free of
subtraction artifacts, and that has predictive and small effects on the shape and contrast of delta function
events. Such a method is demonstrated in \cite{Sub}.

The following additional points were also discussed in paper~I:

{\bf background and variance estimation}: 
The background and variance in real wide-field-of-view
astronomical images cannot be treated as constants over the entire field of view.
Therefore, we suggest to estimate them locally and interpolate.
To estimate the background and variance one needs to make sure that the
estimators are not biased by stars or small galaxies.
Our suggestion is to fit a Gaussian to the histogram of the image pixels
in small regions\footnote{We are currently using $256\times256$~arcsec$^{2}$ blocks.},
and to reject from the fitting process pixels with high
values (e.g., the upper 10\% of pixel values).

{\bf Estimating the transparency}:
The transparency $F_{j}$ of each image
is simply its flux-based photometric zero point.
However, one has to make sure that this zero point is measured
using PSF photometry rather than aperture photometry,
otherwise the zero point may depend also on the seeing.

{\bf Estimating the PSF}:
Among the complications that may affect the PSF measurement
are pixelization and interpolation and the resampling grid.
Furthermore, the PSF is likely not constant spatially and it also
may change with intensity due to charge self repulsion.
This specifically may lead to the brighter-fatter effect (e.g., Walter 2015).
We note that the fact that one needs to estimate the PSF in order to run
this method should not be viewed as a drawback, as any decent method
that finds sources in the image requires this step anyway.

\section{Summary}
\label{sec:Summary}
We present a coaddition method which has the following properties:
\begin{itemize}
\item{It is optimal for the background-noise-dominated case.}
\item{If the input images have equal variance and uncorrelated noise, then the output image has equal variance and uncorrelated noise.}
\item{It provides maximum $S/N$ measurement of the constant sky image on all spatial frequencies.}
\item{It is a sufficient statistic for any hypothesis testing or measurement, and hence represents all the constant-in-time information in the data.}
\item{Its PSF is typically as sharp or sharper than the PSF of the sharpest image in the ensemble.}
\item{It is trivial to program and since it involves only FFT and simple operators, it is also fast to compute.}
\item{Unlike deconvolution or some PSF homogenization techniques, it is numerically stable.}
\item{It provides excellent results for separating binary stars to the diffraction limit in speckle images.}
\item{It applies only local operations on the input images (i.e., convolution with a finite-size PSF). Therefore it allows one to use a slowly-varying position-dependent PSF.}
\end{itemize}
The method can be summarized using two simple equations.
Equation~\ref{eq:Rhat} describes the image coaddition process,
while Equation~\ref{eq:Prhat} provides the PSF of the coadd image.
This method may have far-reaching implications for several fields.
In Figures~\ref{fig:PSFs} through \ref{fig:Binarysprofile} we show how well this methods works on speckle images, and we further discuss this in detail in future papers. 

For seeing-limited surveys like PTF and LSST this method may provide an increase of a few percent to 25\% in the survey speed, compared with popular methods (e.g., \citealp{Annis2014}).
The exact improvement depends on the seeing, background, and transparency distributions.

In addition, it will significantly reduce the need for data access, as all multi-epoch data gathered in the survey could be reduced to a single image, and its PSF, in every sky location. 
Moreover, the images themselves will be easier to process (both in terms of processing speed and memory requirements) because only one image will be needed to be dealt with, instead of many images of the same field (which could amount to thousands of images). 

Another example is the potential of this method for weak lensing studies, that require shape measurements.
An ideal shape measurement on an ensemble of images
is to convolve the model with each of the individual image PSFs, and
simultaneously fit the model to all images with the appropriate weights.
This is a complicated process and as far as we know this is not commonly done.
We showed that fitting a model to our coadd image and using its PSF is equivalent to doing the simultaneous fit on all the images and thus, it is an enabling technique for weak lensing.

%


\acknowledgments

We thank Avishay Gal-Yam, Guy Nir, Ora Zackay, Maayane Soumagnac, Adam Rubin, Frank Masci, Dovi Poznanski and Assaf Horesh for their comments and suggestions.
We thank Guy Nir and Ilan Manulis for help with obtaining the speckle images presented here.
B.Z. is grateful for receiving the Clore fellowship.
E.O.O. is incumbent of
the Arye Dissentshik career development chair and
is grateful for support by
grants from the 
Willner Family Leadership Institute
Ilan Gluzman (Secaucus NJ),
Israel Science Foundation,
Minerva, Weizmann-UK,
and the I-Core program by the Israeli Committee for Planning
and Budgeting and the Israel Science Foundation (ISF).

\end{document}